\documentclass[sigconf]{acmart}

\setcopyright{none} 

\usepackage{url}
\usepackage{mathptmx}
\usepackage{amsmath}
\usepackage{times}
\usepackage{xcolor}
\usepackage{amsmath}
\usepackage[utf8]{inputenc}
\usepackage{array}
\usepackage{multirow}
\usepackage{xspace}
\usepackage{verbatim}
\usepackage{wrapfig}
\usepackage{mathptmx}
\usepackage[english]{babel}
\usepackage{graphicx}
\usepackage{csvsimple}
\usepackage{epstopdf}
\usepackage{datetime}
\usepackage{cleveref}

\usepackage{amssymb}
\usepackage{tabularx}
\usepackage[linesnumbered,ruled,vlined]{algorithm2e}

\newcommand{\projecttitle}{DDPS~}

\begin{document}
\title{Distributed Differential Privacy By Sampling} 

\author{Joshua Joy}

\affiliation{%
    \institution{University of California - Los Angeles}
}
\email{jjoy@cs.ucla.edu}

\begin{abstract}
In this paper, we describe our approach to achieve distributed differential privacy by sampling alone. Our mechanism works in the semi-honest setting (honest-but-curious whereby aggregators attempt to peek at the data though follow the protocol). We show that the utility remains constant and does not degrade due to the variance as compared to the randomized response mechanism. In addition, we show smaller privacy leakage as compared to the randomized response mechanism.
\end{abstract}


\maketitle

\section{Introduction}

Personal mobile information is being continuously collected and analyzed with minimal regard to privacy. As we transition from small mobile personal devices to large-scale sensor collecting self-driving vehicles the needs of privacy increase.

The self-driving vehicle scenario becomes more difficult to preserve both accuracy and privacy due to the smaller number of data owners on the order of hundreds as opposed to large scale deployments of millions of data owners that learn only the heavy hitters (top-k queries).

While many privacy definitions have been proposed, differential privacy has emerged as the gold standard for privacy protection. Differential privacy essentially states that whether or not a single data owner decides to participate in data collection, the final aggregate information will be perturbed only by a negligible amount. That is, the aggregate information released gives no hints to the adversary about a particular data owner. Techniques, such as the Laplace mechanism, add noise calibrated to the sensitivity of the query output~\cite{DBLP:conf/tcc/DworkMNS06}, though do not perturb which data owners actual participate. Thereby making it difficult to privatize graphical structures such as social networks where it is possible to inform data about a targeted data owner from the targets friends alone.

Randomized response has been shown to be optimal in the local privacy setting~\cite{DBLP:conf/nips/DuchiWJ13}. However, in order to preserve accuracy with the randomized response mechanism, privacy must be sacrificed as the data owners must respond truthfully too frequently. For example, a data owner should respond truthfully more than 80\% of the time to have decent accuracy which greatly minimizes any privacy gains~\cite{2016arXiv160404810J,2016arXiv160404892J}. The reason is due to the high variance from the coin tosses. As more aggressive sampling is performed, the variance quickly increases making it difficult to perform accurate estimation of the underlying distribution.

As a result of the accuracy problem, there have been various privacy-preserving systems which focus on the heavy-hitters only~\cite{DBLP:conf/ccs/ErlingssonPK14,DBLP:conf/pet/ChanLSX12}. These techniques ensure privacy only for large populations and can only detect or estimate the most frequently occurring distributions, rather than smaller or less frequently occurring populations.


In this paper, we show how to achieve differential privacy in the distributed setting by sampling alone. We evaluate the accuracy of our privacy-preserving approach utilizing a vehicular crowdsourcing scenario comprising of approximately 50,000 records. In this dataset, each vehicle reports its location utilizing the California Transportation Dataset from magnetic pavement sensors (see Section~\S\ref{sec:evaluation:accuracy}). 
\section{Related Work}

Differential privacy~\cite{DBLP:conf/icalp/Dwork06,DBLP:conf/tcc/DworkMNS06,DBLP:conf/eurocrypt/DworkKMMN06,DBLP:journals/fttcs/DworkR14} has been proposed as a mechanism to privately share data such that anything that can be learned if a particular data owner is included in the database can also be learned if the particular data owner is not included in the database. To achieve this privacy guarantee, differential privacy mandates that only a sublinear number of queries have access to the database and that noise proportional to the global sensitivity of the counting query is added (independent of the number of data owners).

Distributional privacy~\cite{DBLP:journals/jacm/BlumLR13} is a privacy mechanism which says that the released aggregate information only reveals the underlying ground truth distribution and nothing else. This protects individual data owners and is strictly stronger than differential privacy. However, it is computationally inefficient though can work over a large class of queries known as Vapnik-Chervonenkis (VC) dimension.

Zero-knowledge privacy~\cite{DBLP:conf/tcc/GehrkeLP11} is a cryptographically influenced privacy definition that is strictly stronger than differential privacy. Crowd-blending privacy~\cite{DBLP:conf/crypto/GehrkeHLP12} is weaker than differential privacy; however, with a pre-sampling step, satisfies both differential privacy and zero-knowledge privacy. However, these mechanisms do not add noise linear in the number of data owners and rely on aggressive sampling, which negatively impact the accuracy estimations.

The randomized response based mechanisms~\cite{warner1965randomized,fox1986randomized,greenberg1969unrelated,doi:10.1080/01621459.1981.10477741} satisfies the differential privacy mechanism as well as stronger mechanisms such as zero-knowledge privacy. However, the accuracy of the randomized response mechanism quickly degrades unless the coin toss values are configured to large values (e.g., greater than 80\%).
\section{Preliminaries}

\noindent \textbf{Differential Privacy.} Differential privacy has become the \emph{gold standard} privacy mechanism which ensures that the output of a sanitization mechanism does not violate the privacy of any individual inputs.  

\begin{definition}[\cite{DBLP:conf/icalp/Dwork06,DBLP:conf/tcc/DworkMNS06}]{($\epsilon$-Differential Privacy).}
A privacy mechanism $San()$ provides $\epsilon$-differential privacy if, for all datasets $D_1$ and $D_2$ differing on at most one record (i.e., the Hamming distance $H()$ is $H(D_1,D_2) \leq 1$), and for all outputs $O \subseteq Range(San())$:
\begin{equation}
\sup_{D_1,D_2}\frac{\Pr[San(D_1) \in O]}{\Pr[San(D_2) \in O]} \leq exp(\epsilon)
\label{eqn:dp}
\end{equation}
\end{definition}

That is, the probability that a privacy mechanism $San$ produces a given output is almost independent of the presence or absence of any individual record in the dataset.  The closer the distributions are (i.e., smaller $\epsilon$), the stronger the privacy guarantees become and vice versa.

~\\
\noindent \textbf{Private Write.} 
More generally, we assume some class of private information storage~\cite{DBLP:conf/stoc/OstrovskyS97} mechanisms are utilized by the data owner to cryptographically protect their writes to cloud services.

\section{Motivation}

Cryptographic blinding allows a service to compute a function without the service learning the actual input or output. Thus, cryptographic blinding provides private computation. More formally, a client with value $x$ encrypts their value $E(x)$ such that the service can compute $y=f(E(x))$. The client then unblinds $y$ to recover the actual output. The client encryption can be as simple as generating a large random value $r$ and adding to the input $x$ and then subtracting $r$ from $y$.

The randomized response mechanism can be viewed as performing a distributed blinding operation where the privacy is provided by the estimation error that occurs when trying to estimate the distributed blinded value. More formally, let $YES$ refer to the aggregate count of the population that should truthfully respond ``Yes" and $NO$ the population that should truthfully respond ``No" and $s$ is the sampling probability.

Then the randomized response aggregate count is the following $YES_{s_{1}} + (YES+NO)_{s_{2}}=\mathit{AGG}$. The $(YES+NO)_{s_{2}}$ is the distributed blind value that then is estimated by simply calculating the expected sampled value. The final $YES$ value is estimated by subtracting the expected value (of the blind) from the aggregate and then dividing by the sampled probability used for the first part of the term.

It can be seen that the ``distributed blinding" provides a notion of privacy. However, the issue is that the estimation error increases rapidly as the $NO$ population increases due to the standard deviation scaling with the population size.

The remainder of this paper we illustrate our approach which decreases the estimation error and provides strong notions of privacy. The crux of the approach is to provide the ability to remove the error due to the variance by constructing a system of equations.
\section{Distributed Differential Privacy By Sampling}

First, we define the structure of a distributed sampling private mechanism. We then illustrate various mechanisms that satisfy distributed sampling. Finally, we provide the mechanism for preserving accuracy in the distributed privacy model.

\subsection{Structure}


Our mechanism has multiple output value. For our purposes, we use the notation of the $Yes$ population as the ground truth and the remaining data owners are the $No$ population. Each population should blend with each other such that the aggregate information that is released is unable to be used to increase the confidence or inference of an adversary that is trying to determine the value of a specific data owner. 

Data owners are aggressively sampled (e.g., 5\%). To overcome the estimation error due to the large variance, the estimation of the noisy ``Yes" counts and sampled counts are combined to offset each other and effectively cancel the noise, allowing for the aggressive sampling.


\subsection{Sampling}

Sampling whereby a centralized aggregator randomly discards responses has been previously formulated as a mechanism to amplify privacy~\cite{DBLP:conf/crypto/ChaudhuriM06,DBLP:conf/stoc/NissimRS07,DBLP:conf/focs/KasiviswanathanLNRS08,DBLP:conf/ccs/LiQS12,DBLP:conf/crypto/GehrkeHLP12}. The intuition is that when sampling approximates the original aggregate information, an attacker is unable to distinguish when sampling is performed and which data owners are sampled. These privacy mechanisms range from sampling without a sanitization mechanism, sampling to amplify a differentially private mechanism, sampling that tolerates a bias, and even sampling a weaker privacy notion such as k-anonymity to amplify the privacy guarantees. 

However, sampling alone has several issues. First, data owners that answer ``Yes" do not have the protection of strong plausible deniability as they never respond ``No" or are ``forced`` to respond ``Yes" (e.g., via coin tosses). Data owners that answer ``No" do not provide privacy protection as they never answer ``Yes". Second, as we increase the sampling rate the variance will increase rapidly, thus weakening accuracy.  Finally, the privacy strength of the population does not increase as the population increases. The $Yes$ population is fixed (e.g., those at a particular location) and we can only increase the $No$ population. The $No$ population should also contribute noise by answering ``Yes" in order to strengthen privacy.



\subsection{Sampling and Noise}

We could leverage the $No$ population by use the same sampling rate though for the $No$ population have a portion respond ``Yes''. To perform the final estimation we simply subtract the estimated added noise.

~ \\
\noindent \textbf{Sampling and Noise Response.} Each data owner privatizes their actual value $Value$ by performing the following Bernoulli trial. Let $\pi_{s}$ be either the sampling probability for the $Yes$ population as $\pi_{s_{Yes}}$ or for the $No$ population as $\pi_{s_{No}}$.

\begin{equation}
  Privatized~Value =
  \begin{cases}
    \perp & \text{with probability $1-\pi_{s}$} \\
    Value & \text{with probability $\pi_{s}$}
  \end{cases}
\end{equation}

That is, a percentage of the $Yes$ population responds ``Yes" and a percentage of the $No$ population responds ``Yes" (providing noise). However, the $Yes$ data owners do not answer ``No" and also do not have plausible deniability (that is being forced via coin toss to respond ``Yes").

\subsection{Sampling and Plausible Deniability}

We would like to have a percentage of each population respond opposite of their actual value, provide plausible deniability, and have outputs from the space of ``Yes", ``No", and $\perp$ (not participating) in order for the data owners to blend with each other.

To achieve plausible deniability via coin tosses we have a small percentage of the ``Yes" population be ``forced" to respond ``Yes". The other output values follow from the sampling and noise scenario.

\begin{equation}
  Privatized~Value =
  \begin{cases}
    \perp & \text{with probability $1-\pi_{s_{Yes}}$} \\
    1 & \text{with probability } \\
    ~ & \text{$\pi_{s_{Yes}} \times (\pi_1 + (1-\pi_1) \times \pi_2) $} \\
    0 & \text{otherwise}
  \end{cases}
\end{equation}

\begin{equation}
  Privatized~Value =
  \begin{cases}
    \perp & \text{with probability $1-\pi_{s_{No}}$} \\
    1 & \text{with probability} \\
    ~ &  \text{$\pi_{s_{No}} \times ((1-\pi_1) \times \pi_2) $} \\
    0 & \text{otherwise}
  \end{cases}
\end{equation}

A benefit of plausible deniability is that the estimation of the population will provides privacy protection via noise. However, it is difficult to estimate the underlying ground truth due to the added noise. We desire better calibration over the privacy mechanism.

\subsection{Mechanism}


The output mechanism has multiple output values. A fraction of each of the $Yes$ and $No$ population should be included. Each population should have some notion of plausible deniability. The total number of data owners $DO$ can be computed by summing the total number of ``Yes", ``No", and $\perp$ responses. It should be noted that if the data owners would not write $\perp$ it would be difficult to estimate and calculate the underlying $Yes$ count.

\begin{definition}(Minimal Variance Parameters)
We model the sum of independent Bernoulli trials that are not identically distributed, as the poisson binomial distribution (we combine ``No" and $\perp$ into the same output space for modeling purposes). Let $Yes'$ represent those that respond ``Yes", regardless if they are from the $Yes$ or $No$ population. 

The success probabilities are due to the contributions from the $Yes$ and $No$ populations. We then sum and search for the minimum variance.

Utility is maximized when:

\begin{align}
\label{alg:minvariance}
\begin{split} 
\min({Var(P(``Yes"|Yes))+Var(P(``Yes"|No))})
\end{split}
\end{align}
\end{definition}

There are a couple observations. The first is that uniform sampling across both populations ($Yes$ and $No$) limits the ability to achieve optimal variance. As we increase the $No$ population by increasing the queries and the number of data owners that participate, the variance will correspondingly increase. For example, 10\% sampling will incur a large variance for a population of one million data owners. To address this, the sampling parameters should be separately tuned for each population. We desire a small amount of data owners to be sampled from the $Yes$ population to protect privacy and an even smaller amount from the $No$ population (as this population will be large and only a small amount is required for linear noise). The other observation is that for the plausible deniability, by fixing the probabilities the same across the $Yes$ and $No$ population also restricts the variance that can be achieved.

Thus, the optimal distributed sampling mechanism is one that tunes both populations.

~\\
\noindent \textbf{ Mechanism.} \label{sec:optimalmechanism} Let $\pi_{s}$ be either the sampling probability for the $Yes$ population as $\pi_{s_{Yes}}$ or for the $No$ population as $\pi_{s_{No}}$. Let $\pi_p$ be either the plausible deniability parameter for the $Yes$ population as $\pi_1$ or the $No$ population $\pi_2$ respectively. The  mechanism that we use which satisfies distributed sampling is as follows:

\begin{equation}
  Privatized~Value =
  \begin{cases}
    \perp & \text{with probability} \\
    ~ & \text{$1-(\pi_{s_{Yes_{1}}}+\pi_{s_{Yes_{2}}})$} \\
    1 & \text{with probability $\pi_{s_{Yes_{1}}} \times \pi_1$} \\
    1 & \text{with probability $\pi_{s_{Yes_{2}}} \times \pi_2$} \\
    0 & \text{with probability} \\
    ~ & \textbf{$\pi_{s_{Yes_{1}}} \times (1 - \pi_1) +$} \\
    ~ & \textbf{$\pi_{s_{Yes_{2}}} \times (1 - \pi_2)$} 
  \end{cases}
\end{equation}

\begin{equation}
  Privatized~Value =
  \begin{cases}
    \perp & \text{with probability $1-\pi_{s_{No}}$} \\
    1 & \text{with probability} \\
    ~ &  \text{$\pi_{s_{No}} \times \pi_3 $} \\
    0 & \text{otherwise}
  \end{cases}
\end{equation}

It should be noted that \textit{$\pi_{s_{Yes_{1}}} \times \pi_1$} are the percentage of data owners that answer truthfully ``Yes" and \textit{$\pi_{s_{Yes_{2}}} \times \pi_2$} are the percentage of data owners that are ``forced" to respond ``Yes" providing the plausible deniability. Each case has its own coin toss parameters in order to be able to fine tune the variance and reduce the estimation error as opposed to the prior examples where the variance cascades across terms adding error.

\stepcounter{equation}
\stepcounter{equation}

\section{Accuracy}
\label{sec:accuracy}

\subsection{First Attempt Cancelling the Noise}

Let $DO$ be the total number of data owners. Let $YES$ and $NO$ be the population count of those that truthfully respond ``Yes" and ``No" respectively such that $YES + NO=DO$.


\begin{lemma}(Yes Estimate From Aggregated Count)

~\\
Expected value of ``Yes" responses is:

\begin{align}
\begin{split}
E[``Yes"] & = \pi_{Yes_{1}} \times \pi_1 \times \mathit{YES} + \\
& \pi_{Yes_{2}} \times \pi_2 \times \mathit{YES} + \\
& \pi_{No} \times \pi_3 \times (\mathit{DO}-\mathit{YES})
\end{split}
\end{align}

Solving for $YES$ results in:

\begin{align}
\begin{split}
\mathit{YES} & = \frac{E[``Yes"] - (\pi_{No} \times \pi_3 \times \mathit{DO})}{(\pi_{Yes_{1}} \times \pi_1) + (\pi_{Yes_{2}} \times \pi_2) - (\pi_{No} \times \pi_3 )}
\end{split}
\end{align}

The estimator $\mathit{\hat{YES}_{Yes}}$ accounting for the standard deviation $\sigma(``Yes")$ is:

\begin{align}
\begin{split}
\mathit{\hat{YES_{Yes}}} & = \frac{E[``Yes"] \pm \sigma(``Yes") - (\pi_{No} \times \pi_3 \times \mathit{DO})}{(\pi_{Yes_{1}} \times \pi_1) + (\pi_{Yes_{2}} \times \pi_2) - (\pi_{No} \times \pi_3 )}
\end{split}
\end{align}

\end{lemma}

\begin{lemma}{Standard Deviation of the Aggregated ``Yes" Count}

The standard deviation $\sigma(``Yes")$ is:

\begin{align}
\begin{split}
Var(``Yes") & =  ((\pi_{Yes_{1}} \times \pi_1 + \pi_{Yes_{2}} \times \pi_2) \times \\
& (1 - (\pi_{Yes_{1}} \times \pi_1 + \pi_{Yes_{2}} \times \pi_2)) \times \\
& \mathit{YES}) + \\
& (\pi_{No} \times \pi_3 \times \\
& (1 - (\pi_{No} \times \pi_3)) \times \\
& \mathit{NO} )
\end{split}
\end{align}

\begin{align}
\begin{split}
\sigma(``Yes") &= \sqrt{Var(``Yes")}
\end{split}
\end{align}

\end{lemma}


\begin{lemma}(Yes Estimate From Aggregated ``No" Count)

~\\
Expected value of ``Yes" responses is:

\begin{align}
\begin{split}
E[``Yes"] & = \pi_{Yes_{1}} \times (1-\pi_1) \times \mathit{YES} + \\
& \pi_{Yes_{2}} \times (1-\pi_2) \times \mathit{YES} + \\
& \pi_{No} \times (1-\pi_3) \times (\mathit{DO}-\mathit{YES})
\end{split}
\end{align}

Solving for $YES$ results in:

\begin{align}
\begin{split}
\mathit{YES} & = \frac{E[``Yes"] - (\pi_{No} \times (1-\pi_3) \times \mathit{DO})}{(\pi_{Yes_{1}} \times (1-\pi_1)) + (\pi_{Yes_{2}} \times (1-\pi_2)) - (\pi_{No} \times (1-\pi_3) )}
\end{split}
\end{align}

The estimator $\mathit{\hat{YES}_{No}}$ accounting for the standard deviation $\sigma(``No")$ is:

\begin{align}
\begin{split}
\mathit{\hat{YES_{No}}} & = \frac{E[``Yes"] \pm \sigma(``Yes") - (\pi_{No} \times (1-\pi_3) \times \mathit{DO})}{(\pi_{Yes_{1}} \times (1-\pi_1)) + (\pi_{Yes_{2}} \times (1-\pi_2)) - (\pi_{No} \times (1-\pi_3) )}
\end{split}
\end{align}

\end{lemma}

\begin{lemma}{Standard Deviation of the Aggregated ``No" Count}

The standard deviation $\sigma(``No")$ is:

\begin{align}
\begin{split}
Var(``Yes") & =  ((\pi_{Yes_{1}} \times (1-\pi_1) + \pi_{Yes_{2}} \times (1-\pi_2)) \times \\
& (1 - (\pi_{Yes_{1}} \times (1-\pi_1) + \pi_{Yes_{2}} \times (1-\pi_2))) \times \\
& \mathit{YES}) + \\
& (\pi_{No} \times (1-\pi_3) \times \\
& (1 - (\pi_{No} \times (1-\pi_3))) \times \\
& \mathit{NO} )
\end{split}
\end{align}

\begin{align}
\begin{split}
\sigma(``No") &= \sqrt{Var(``No")}
\end{split}
\end{align}

\end{lemma}


\begin{lemma}(Yes Estimate From Sampled Population)
~\\
Expected value of $\perp$ (not participating) responses is:

\begin{align}
\begin{split}
E[\perp] & = (1 - (\pi_{Yes_{1}} + \pi_{Yes_{2}})) \times \mathit{YES} + \\
& (1 - \pi_{No}) \times (\mathit{DO} - \mathit{YES})
\end{split}
\end{align}

Solving for $YES$ results in:

\begin{align}
\begin{split}
\mathit{YES} & = \frac{E[\perp] - ((1 - \pi_{No}) \times \mathit{DO})}{(1 - (\pi_{Yes_{1}} + \pi_{Yes_{2}})) - (1 - \pi_{No})}
\end{split}
\end{align}

The estimator $\mathit{\hat{YES}_{\perp}}$ accounting for the standard deviation $\sigma(\perp)$ is:

\begin{align}
\begin{split}
\mathit{\hat{YES_{\perp}}} & = \frac{E[\perp] \pm \sigma(\perp) - ((1 - \pi_{No}) \times \mathit{DO})}{(1 - (\pi_{Yes_{1}} + \pi_{Yes_{2}})) - (1 - \pi_{No})}
\end{split}
\end{align}

\end{lemma}

\begin{lemma}{Standard Deviation of the Sampled Population}

The standard deviation $\sigma(\perp)$ is:

\begin{align}
\begin{split}
Var(\perp) & =  ((1 - (\pi_{Yes_{1}} + \pi_{Yes_{2}})) \times \\
& (\pi_{Yes_{1}} + \pi_{Yes_{2}}) \times \\
& \mathit{YES}) + \\
& ((1 - \pi_{No}) \times \\
& \pi_{No} \times \\
& \mathit{NO} )
\end{split}
\end{align}

\begin{align}
\begin{split}
\sigma(\perp) &= \sqrt{Var(\perp)}
\end{split}
\end{align}

\end{lemma}


\begin{lemma}{Solving for $YES$}

There are two approaches we can take. We can either use the ``Yes" estimators to estimate the underlying population as described earlier. Or we can treat the equations as a system of linear equations.

The observation is that setting $\pi_1=\pi_2=\pi_3=1$ results in the standard deviation being equal for $\sigma(``Yes")$, $\sigma(``No")$, and $\sigma(\perp)$. This has the effect of resulting in no ``No" responses and the two equations are thus dependant.

We have the following system of linear equations of two unknown variables $YES$ and $\sigma$ as follows:

\begin{align}
E[``Yes"] \pm \sigma = Observed(``Yes") \\
E[\perp] \pm \sigma = Observed(\perp) \\
E[``Yes"] \pm \sigma + E[\perp] \pm \sigma = DO
\end{align}

We then solve for $YES$ and $\sigma$ for each combination of varying $\pm$ signs using a solver. We eliminate the solutions which assign $YES$ a negative value.



\end{lemma}

It would be nice if we could cancel out the error. It would also be nice if the system of linear equations above would have exactly one solution. However, it's not clear that we can immediately guarantee this.

\subsection{Reducing the Error}

As we control the randomization, can we construct a mechanism whereby the error introduced by the $NO$ population cancels out? Performing uniform sampling across both $YES$ and $NO$ populations allows us to cancel the \textit{population} error though we are not able to precisely estimate the $YES$ population as it also cancels out the $YES$ terms. 

One observation is that the error terms potentially could cancel out if the signs were flipped. Thus, we construct our mechanism as follows. Each data owner responds \textit{twice} for the same query, though slightly flips a single term to allow for the error cancelation.

\begin{equation}
  \mathit{A}~Privatized~Value =
  \begin{cases}
    \perp_1 & \text{with probability $\pi_{\perp_1}$} \\
    \perp_1 & \text{with probability $\pi_{s}$} \\
    \perp_2 & \text{with probability $\pi_{\perp_2}$} \\
    \perp_3 & \text{with probability} \\
    		& \text{$1-\pi_{\perp_1}-\pi_{\perp_2}-\pi_{s}$}
\end{cases}
\end{equation}

\begin{equation}
  \mathit{B}~Privatized~Value =
  \begin{cases}
    \perp_1 & \text{with probability $\pi_{\perp_1}$} \\
    \perp_2 & \text{with probability $\pi_{s}$} \\
    \perp_2 & \text{with probability $\pi_{\perp_2}$} \\
    \perp_3 & \text{with probability} \\
    		& \text{$1-\pi_{\perp_1}-\pi_{\perp_2}-\pi_{s}$}
\end{cases}
\end{equation}

\begin{equation}
  \mathit{YES_C}~Privatized~Value =
  \begin{cases}
    \perp_1 & \text{with probability $\pi_{\perp_1}$} \\
    \perp_2 & \text{with probability $\pi_{s}$} \\
    \perp_2 & \text{with probability $\pi_{\perp_2}$} \\
    \perp_3 & \text{with probability} \\
                 & \text{$1-\pi_{\perp_1}-\pi_{\perp_2}-\pi_{s}$}
\end{cases}
\end{equation}

\begin{equation}
  \mathit{NO_C}~Privatized~Value =
  \begin{cases}
    \perp_1 & \text{with probability $\pi_{\perp_1}$} \\
    \perp_3 & \text{with probability $\pi_{s}$} \\
    \perp_2 & \text{with probability $\pi_{\perp_2}$} \\
    \perp_3 & \text{with probability} \\
    		& \text{$1-\pi_{\perp_1}-\pi_{\perp_2}-\pi_{s}$}
\end{cases}
\end{equation}

The expected values are as follows.

The first set expected values are:

\begin{align}
\begin{split}
E[\perp_{1_{A}}] & = \pi_{\perp_1} \times \mathit{TOTAL} + \pi_{s} \times \mathit{TOTAL} \\
E[\perp_{2_{A}}] & = \pi_{\perp_2} \times \mathit{TOTAL} \\
E[\perp_{3_{A}}] & = \pi_{\perp_3} \times \mathit{TOTAL} 
\end{split}
\end{align}

The second set expected values are:

\begin{align}
\begin{split}
E[\perp_{1_{B}}] & = \pi_{\perp_1} \times \mathit{TOTAL}  \\
E[\perp_{2_{B}}] & = \pi_{\perp_2} \times \mathit{TOTAL} + \pi_{s} \times \mathit{TOTAL} \\
E[\perp_{3_{B}}] & = \pi_{\perp_3} \times \mathit{TOTAL} 
\end{split}
\end{align}

The third set expected values are:

\begin{align}
\begin{split}
E[\perp_{1_{C}}] & = \pi_{\perp_1} \times \mathit{TOTAL}  \\
E[\perp_{2_{C}}] & = \pi_{\perp_2} \times \mathit{TOTAL} + \pi_{s} \times \mathit{YES} \\
E[\perp_{3_{C}}] & = \pi_{\perp_3} \times \mathit{TOTAL} + \pi_{s} \times \mathit{NO}
\end{split}
\end{align}

To solve for the \textit{YES} population as follows:

\begin{align}
\begin{split}
\mathit{YES} & = \frac{E[\perp_{2_{C}}] - E[\perp_{2_{A}}]}{\pi_{s}}
\end{split}
\end{align}

\subsection{Multiple Values}

We now examine how to privatize for multiple for multiple values. We can achieve this in two rounds. Say there are $V$ possible output values (e.g., V possible location IDs). Let $V'$ be the truthful output value.

\begin{equation}
  Round~One =
  \begin{cases}
    \perp_0 & \text{with probability $\pi_{\perp_0}$} \\
    \perp_0 & \text{with probability $\pi_{s}$} \\
    \perp_1 & \text{with probability $\pi_{\perp_1}$} \\
    \perp_2 & \text{with probability $\pi_{\perp_2}$} \\
    ... \\
    \perp_V & \text{with probability} \\
    		& \text{$1-\pi_{\perp_0}-\pi_{\perp_1}-...-\pi_{\perp_V}-\pi_{s}$}
\end{cases}
\end{equation}

\begin{equation}
  Yes~Round~Two =
  \begin{cases}
    \perp_0 & \text{with probability $\pi_{\perp_0}$} \\
    \perp_1 & \text{with probability $\pi_{\perp_1}$} \\
    \perp_2 & \text{with probability $\pi_{\perp_2}$} \\
    ... \\
    \perp_{V'} & \text{with probability $\pi_{v'}$} \\
    \perp_{V'} & \text{with probability $\pi_{s}$} \\
    ... \\
    \perp_V & \text{with probability} \\
    		& \text{$1-\pi_{\perp_0}-\pi_{\perp_1}-...-\pi_{\perp_V}-\pi_{s}$}
\end{cases}
\end{equation}

\begin{equation}
  No~Round~Two =
  \begin{cases}
    \perp_0 & \text{with probability $\pi_{\perp_0}$} \\
    \perp_0 & \text{with probability $\pi_{s}$} \\
    \perp_1 & \text{with probability $\pi_{\perp_1}$} \\
    \perp_2 & \text{with probability $\pi_{\perp_2}$} \\
    ... \\
    \perp_V & \text{with probability} \\
    		& \text{$1-\pi_{\perp_0}-\pi_{\perp_1}-...-\pi_{\perp_V}-\pi_{s}$}
\end{cases}
\end{equation}

The expected values are as follows.

The first round of expected values are:

\begin{align}
\begin{split}
E[\perp_{0_{0}}] & = \pi_{\perp_0} \times \mathit{TOTAL} + \pi_{s} \times \mathit{TOTAL} \\
E[\perp_{1_{0}}] & = \pi_{\perp_1} \times \mathit{TOTAL} \\
E[\perp_{2_{0}}] & = \pi_{\perp_2} \times \mathit{TOTAL} \\
... \\
E[\perp_{V_{0}}] & = \pi_{\perp_V} \times \mathit{TOTAL}
\end{split}
\end{align}

The second round of expected values are:

\begin{align}
\begin{split}
E[\perp_{0_{1}}] & = \pi_{\perp_0} \times \mathit{TOTAL} \\
E[\perp_{1_{1}}] & = \pi_{\perp_1} \times \mathit{TOTAL} \\
E[\perp_{2_{1}}] & = \pi_{\perp_2} \times \mathit{TOTAL} \\
... \\
E[\perp_{V'_{1}}] & = \pi_{\perp_{V'}} \times \mathit{TOTAL} + \pi_{s} \times \mathit{TOTAL} \\
... \\
E[\perp_{V_{1}}] & = \pi_{\perp_V} \times \mathit{TOTAL}
\end{split}
\end{align}

To solve for the \textit{YES} population as follows:

\begin{align}
\begin{split}
\mathit{YES} & = \frac{E[\perp_{V'_{1}}] - E[\perp_{V'_{0}}]}{\pi_{s}}
\end{split}
\end{align}

\section{Privacy Guarantee}

\subsection{Differential Privacy Guarantee}

\projecttitle satisfies differential privacy. Let ``Yes'' be the privatized response and $Yes$ and $No$ be the respective populations.


\begin{equation}
\epsilon_{DP} = \max \Bigg({\ln \bigg(\frac{\Pr[\pi_{\perp_2} | YES]} {\Pr[\pi_{\perp_2} | NO]} \bigg),\ln \bigg(\frac{\Pr[\pi_{\perp_3} | NO]} {\Pr[\pi_{\perp_3} | YES]} \bigg)} \Bigg)
\end{equation}

\begin{equation}
\frac{\Pr[\pi_{\perp_2} | YES]} {\Pr[\pi_{\perp_2} | NO]} = \frac{ \pi_{\perp_2} + \pi_{s}}{ \pi_{\perp_2} }
\end{equation}

\begin{equation}
\frac{\Pr[\pi_{\perp_3} | NO]} {\Pr[\pi_{\perp_3} | YES]}  = \frac{ \pi_{\perp_3} + \pi_{s} }{ \pi_{\perp_3} }
\end{equation}

\begin{equation}
\epsilon_{DP} = \max \Bigg( \frac{ \pi_{\perp_2} + \pi_{s}}{ \pi_{\perp_2} },\frac{ \pi_{\perp_3} + \pi_{s} }{ \pi_{\perp_3} } \Bigg) 
\end{equation}

\section{Evaluation}

\begin{figure*}[ht!]
    \includegraphics[width=1\textwidth]{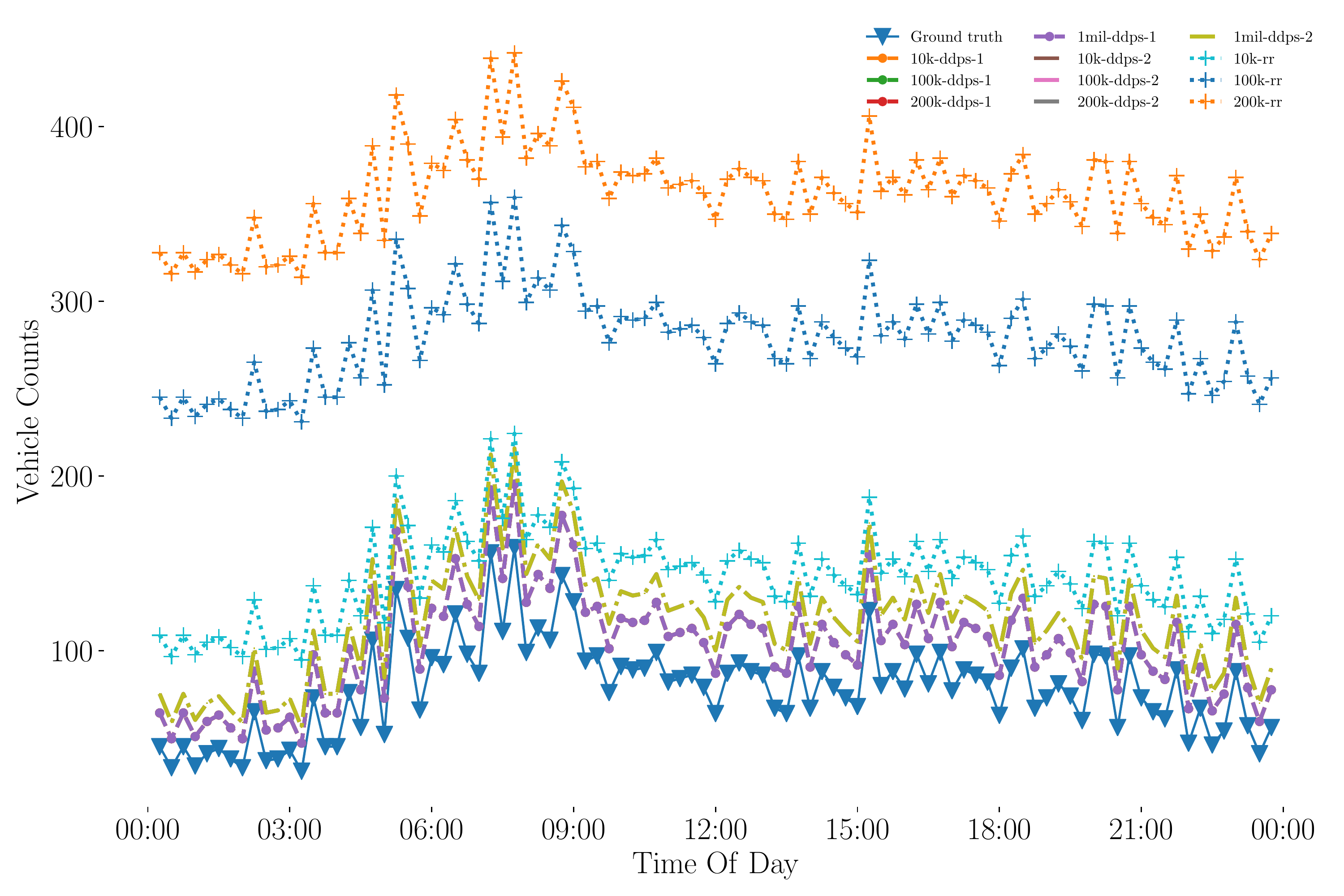}
    \caption{\textbf{Accuracy.} Ground truth versus privatized vehicle counts with a 99\% confidence interval. The population not at the station being monitored (i.e., $No$ population) is increases by expanding the query to the population listed in the table.}
        \label{fig:trafficdata}
\end{figure*}

\subsection{Accuracy}
\label{sec:evaluation:accuracy}

We evaluate the Distributed Differential Privacy By Sampling (DDPS) mechanism over a real dataset rather than arbitrary distributions. We utilize the California Transportation Dataset from magnetic pavement sensors~\cite{pems} collected in LA$\backslash$Ventura California freeways~\cite{cwwpinformation}.  There are a total of 3,220 stations and 47,719 vehicles total. We assign virtual identities to each vehicle. Each vehicle announces the station it is currently at.

Figure~\ref{fig:trafficdata} compares the DDPS mechanism to the ground truth data over a 24 hour time period with a confidence interval of 99\%. We select a single popular highway station that collects and aggregates vehicle counts every 30 seconds. (For reasons of illustration we graph a subset of points for readability). We assign virtual identifiers and have every vehicle at the monitored station truthfully report ``Yes" while every other vehicle in the population truthfully reports ``No". The \projecttitle mechanism then privatizes each vehicle's response. Traffic management analyzing the privatized time series would be able to infer the ebbs and flow of the vehicular traffic. 

The coin toss probabilities are fixed and the sampling $No$ rate is adjusted for privacy and error guarantees. The parameters $\pi_{s_{Yes_{1}}}=0.45$, $\pi_{s_{Yes_{2}}}=0.50$,  $\pi_1=0.95$, $\pi_2=0.98$, $\pi_3=0.98$ are fixed and $\pi_{s_{No}}=0.068$ varies from $0.068$ to $0.00000068$.

\subsection{Privacy}
\label{sec:evaluation:privacy}

\begin{figure*}[ht!]
    \includegraphics[width=1\textwidth]{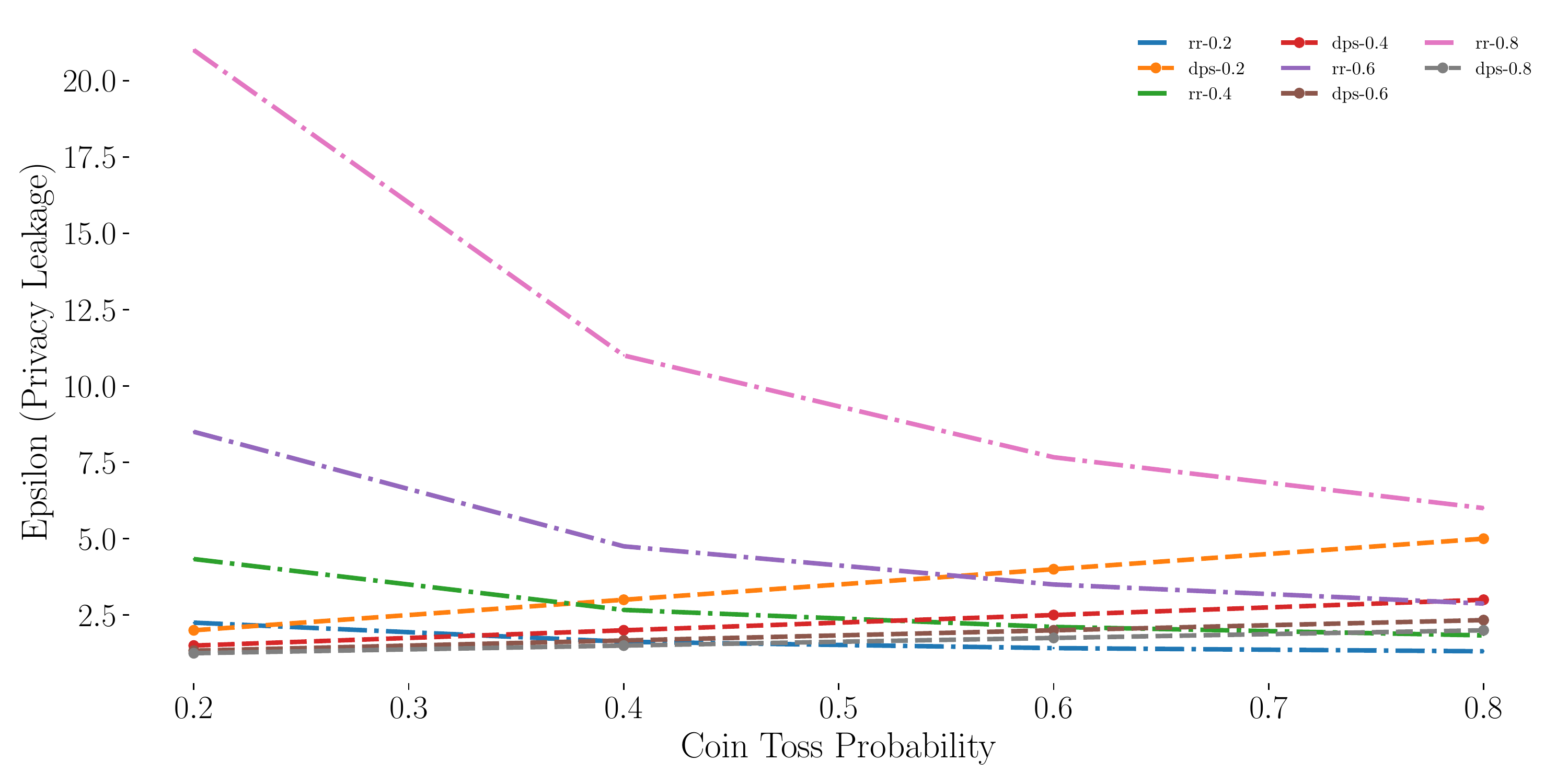}
    \caption{\textbf{Privacy Leakage.} Differential Privacy $\epsilon$ leakage.}
        \label{fig:epsilon}
\end{figure*}

\begin{figure}[ht!]
    \includegraphics[width=1\columnwidth]{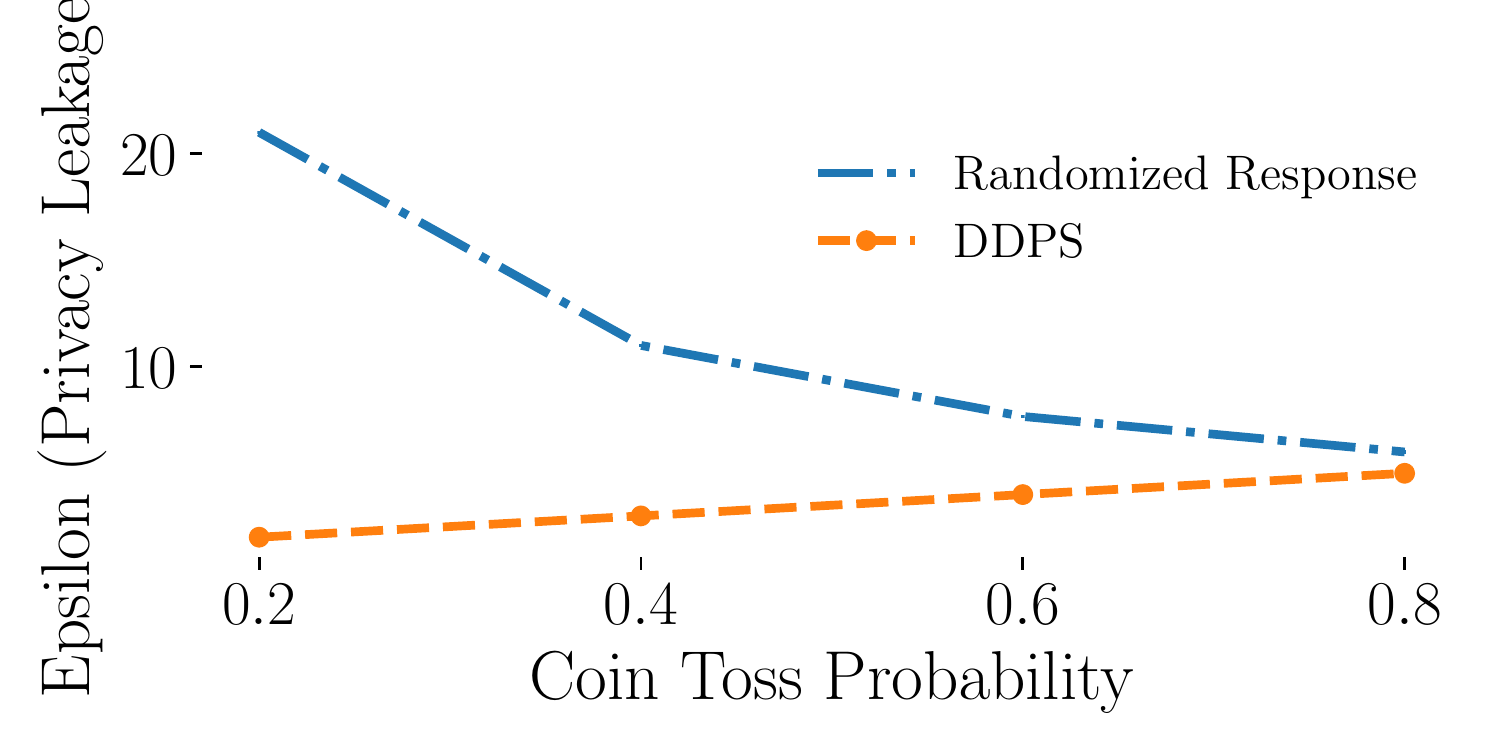}
    \caption{\textbf{Privacy Leakage.} Differential Privacy $\epsilon$ leakage.}
        \label{fig:epsilon}
\end{figure}

We evaluate privacy strength of the \projecttitle mechanism by examining the crowd sizes. We use the binomial complementary cumulative distribution function (ccdf) to determine the threshold of the number of successes. The crowd blending size is controlled by the $No$ population, that is the number of data owners that truthfully respond ``No". Also, the data owner blends in the $\perp$ crowd, that is the crowd which is not participating.


The number of crowds an single owner blends is determined by the number of queries (e.g., number of crowdsourced locations). The coin toss probability of  ($\pi_{s_{No}} \times \pi_3$) magnifies the number of crowds, that is the number of queries that the data owner will respond to. Also, the data owner blends in the $\perp$ crowd.

Table~\ref{tab:locations} shows the number of additional locations a single data owner will respond to out of 3,320. Decreasing the sampling rate causes a reduction in the number of additional locations. In this case, the mechanism essentially ignores the data owner as the aggregate output information is essentially the same output distribution as show by the graph. As each data owner reports they are at their actual location less than 5\% of the time, the additional three simultaneous announcements provides adequate privacy protection.

Table~\ref{tab:dataowners} shows the number of data owners an individual will blend with. We vary the population of additional data owners and accordingly decrease the sampling rate $\pi_{No}$ to calibrate the error. The noise added is linear with the number of data owners to protect. Sampling reduces the the crowd size yet at the same time provides privacy protection.

\begin{table}[]
\Huge
\centering
\resizebox{\columnwidth}{!}{%
\begin{tabular}{l|l|l|l}
\textbf{Locations} & \textbf{TossThree($\pi_3$)} & \textbf{SamplingRate($\pi_{s_{No}}$)} & \textbf{Number Stations} \\ \hline
1                & 0.98              & 0.05  & 3320        \\ \hline
3                & 0.98             & 0.00025   & 83,000        \\ \hline              
3                & 0.98             & 0.000025   & 830,000        \\ \hline                          
\end{tabular}%
}
\caption{\textbf{Multiple Locations.} A single data owner will claim to be in multiple locations simultaneously.}
\label{tab:locations}
\end{table}

\begin{table}[]
\Huge
\centering
\resizebox{\columnwidth}{!}{%
\begin{tabular}{l|l|l|l}
\textbf{CrowdSize} & \textbf{TossThree($\pi_3$)} & \textbf{SamplingRate($\pi_{s_{No}}$)} & \textbf{Population} \\ \hline
160                   & 0.98              & 0.05       &  48,719           \\ \hline
140                  & 0.98             & 0.00025        & 1,047,719            \\ \hline
130                  & 0.98             & 0.000025        & 10,047,719            \\ \hline           
\end{tabular}%
}
\caption{\textbf{Crowd size.} A single data owner blends with $k$ other data owners who noisily answer ``Yes" from the $No$ population .}
\label{tab:dataowners}
\end{table}

\section{Conclusion}

In this paper, we show how to achieve differential privacy in the distributed setting by sampling alone. We show that we can maintain constant error (i.e., absolute error) even as the population increases and achieve lower privacy leakage as compared to randomized response.



\bibliographystyle{ACM-Reference-Format}
\bibliography{fss,learning,localprivacy,main,mpc,privacy,streamprivacy,vehicles}

\end{document}